# Physics-informed machine learning improves detection of head impacts

Samuel J. Raymond[1,#], Nicholas J. Cecchi[1], Hossein Vahid Alizadeh[1], Ashlyn A. Callan[1], Eli Rice[2], Yuzhe Liu[1], Zhou Zhou[1], Michael Zeineh[3], David B. Camarillo[1,4,5]

[1] Department of Bioengineering, Stanford University, Stanford, CA, 94305, USA.

[2] Stanford Center for Clinical Research, Stanford University, Stanford, CA, 94305, USA.

[3] Department of Radiology, Stanford University, Stanford, CA, 94305, USA.

[4] Department of Neurosurgery, Stanford University, Stanford, CA, 94305, USA.

[5] Department of Mechanical Engineering, Stanford University, Stanford, CA, 94305, USA.

[#] Corresponding author (e-mail: sjray@stanford.edu)

## Abstract

In this work we present a new physics-informed machine learning model that can be used to analyze kinematic data from an instrumented mouthguard and detect impacts to the head. Monitoring player impacts is vitally important to understanding and protecting from injuries like concussion. Typically, to analyze this data, a combination of video analysis and sensor data is used to ascertain the recorded events are true impacts and not false positives. In fact, due to the nature of using wearable devices in sports, false positives vastly outnumber the true positives. Yet, manual video analysis is time-consuming. This imbalance leads traditional machine learning approaches to exhibit poor performance in both detecting true positives and preventing false negatives. Here, we show that by simulating head impacts numerically using a standard Finite Element head-neck model, a large dataset of synthetic impacts can be created to augment the gathered, verified, impact data from mouthguards. This combined physics-informed machine learning impact detector reported improved performance on test datasets compared to traditional impact detectors with negative predictive value and positive predictive values of 88% and 87% respectively. Consequently, this model reported the best results to date for an impact detection algorithm for American Football, achieving an $F_1$ score of 0.95. In addition, this physics-informed machine learning impact detector was able to accurately detect true and false impacts from a test dataset at a rate of 90% and 100% relative to a purely manual video analysis workflow. Saving over 12 hours of manual video analysis for a modest dataset, at an overall accuracy of 92%, these results indicate that this model could be used in place of, or alongside, traditional video analysis to allow for larger scale and more efficient impact detection in sports such as American Football.

**Keywords** instrumented mouthguard, traumatic brain injury, American football, concussion, deep learning, physics-informed machine learning

## 1. Introduction

Collisions are commonplace in contact sports such as American Football. Despite improvements in protective equipment (shoulder pads, helmets, etc.), the risk of head injuries remains [1,2]. Threats of concussion and mild traumatic brain injuries (mTBIs) represent serious health consequences for these athletes [3]. During games and practice sessions, players' heads may be subjected to accelerations that result in high levels of internal strains within the brain [4] and can result in physiological damage. Left untreated, mTBI can contribute to neurodegenerative diseases after repeated exposure to impacts [5]. Due to its mild nature, mTBI is difficult to detect and players are often unaware of the magnitude of the impact they sustain. This makes mTBI-causing impacts virtually impossible to observe during a game or practice session based solely on real-time or video analysis [6], unless symptoms are immediately present. One solution to this problem involves attaching sensors to players and recording their kinematics during gameplay to measure the impacts their body receives [7], and verifying this with video analysis. Instrumented mouthguards, in particular, have been used in the past to gather data on athlete head kinematics and develop models of the risk of brain injury [8,9]. Mouthguards are well suited to track head motion as they are fitted to the upper dentition, and therefore directly connected to the skull [10]. With this additional source of data, combined with video analysis and player interviews, the study of head impacts, concussion, and mTBI has resulted in a number of key insights to help improve helmet design and safety [11]. However, in attempts to both improve the accuracy of these instrumented mouthguards and to alleviate the time-consuming procedure of video analysis, impact detection algorithms have also been developed [12–14]. The goal of these algorithms is to complement and/or remove the need for manual video analysis and can be used to identify when a real impact has occurred either on or off screen, and to differentiate between true and false impacts. However, many unrelated events such as chewing, improper placement of mouthguard (such as in the sock during running), and device malfunctions can result in false impact signals[15]. While signal filtering techniques [9] are used to reduce the number of these false impacts, oftentimes these signals are too complex and similar to differential by standard means. To overcome these limitations, new impact detection techniques have been developed based on machine learning tools such as support vector machines (SVM) [16] and deep neural networks (DNNs)[13,17,18]. These models are trained on the video verified data of true and false impacts to either use a complex set of features (in the case of SVM models) or learn a set of features from the data (in the case of DNNs) to best identify true impact signals from the false impact signals based on the kinematic data. This reliance on verified field data, however, represents a limitation for these machine learning models since, in practice, there are far more false impacts than true impacts, and the data that can be collected is limited by the time-consuming work that is involved in video analysis. A potential source of new data that can be leveraged for training these models, could be from the developments in the last decade of Finite Element Head-Neck models [17,18] and the data these and more complex, biomechanical models can provide.

Despite the promising advance in neuroimaging techniques, the clinical diagnosis of brain injuries remains challenging. This is particularly true for mTBI since the brain often exhibits no identifiable structural alteration based on conventional neuroimaging techniques (e.g., computed tomography and magnetic resonance imaging) [19]. As numerical surrogates, finite element models of the human head can play an important role to narrow down the mechanobiological gap between the mechanical force during an impact and the resulting brain injury [20–23]. These mathematical models, which incorporate an appropriate anatomical representation of the head, sophisticated material properties of relevant brain structures as well as precise descriptions of interfacial among various intracranial components, offer spatiotemporal detailed information of the tissue responses. This information is often interpreted by means of injury metrics (e.g., maximum principal strain, maximum tract-oriented strain, and strain rate), informing not only the severity and risk of brain injury from impacts during sports such as American football [24], but also the development of head protection strategies such as helmet design [1,25]. However, the reliability of inferences drawn from these models depends heavily on the accuracy of the kinematic inputs. Using data from an instrumented mouthguard is one very useful approach [9]. Another way to generate useful data is to subject these head models to scenarios that represent the environment of a real head impact. By using a physical dummy model of the head and generating a landscape of impact scenarios, an endless source of additional kinematic data becomes available.

Combining simulation data and real-world data for improved machine learning is a useful technique that has gained in adoption since the rise of physics-informed machine learning (PIML) [26–34]. PIML is a cutting-edge new field that sits at the intersection of scientific computing and machine learning. During the immense rise to power of Artificial Intelligence and Machine Learning over the last two decades, scientific computing was not largely looked at as a field of direct application. As a fundamentally probabilistic endeavor, deep learning may infer values that might violate or exceed reasonable bounds, which would violate the Laws used in numerical physics, potentially rendering it fundamentally flawed as an application. Scientific data and numerical models, however, are themselves not perfect, with errors and systematic biases present in these approaches as well [32–34]. PIML is a broad field that encompasses applications such as physics-informed neural networks [28,33], learning equations from data [30,31] and utilizing synthetic data for training and testing [26,27]. Synthetic data can come from simplistic numerical models based on empirical formulae or can be produced from state-of-the-art computational mechanics simulators [18,35–41]. In this work the Finite Element head model will be subjected to a range of impact conditions to generate kinematic data of the same form of data produced by the instrumented mouthguard worn by players as they undergo similar impacts on the field. We will show that by including more synthetic data representing true impacts to our DNN impact detector, our algorithm will perform better than one relying only on field data, and show that this PIML-based impact detector can be used as a replacement for traditional video analysis, potentially saving hundreds of hours of manual labor. To do so, this paper is structured as follows: the next section will explain the methods used to gather the field data, and generate the synthetic data from simulations, as well as the strategy used to improve the performance of the impact detector on test data. Next the results section will detail the improvements of the algorithm as a function of the amount of synthetic data used to train with, and report a direct comparison between a fully manual workflow and a fully automated workflow. Finally, the discussion section will cover the pros and cons of this approach, its limitations, and the implications that this new PIML-based impact detector may have as a replacement to traditional video analysis workflows.

## 2. Methods

In this section we describe the procedures used to gather the on-field data used to train our impact detector. The Finite Element head-neck model is also introduced as well as the model parameters used to generate the synthetic data. Finally, the strategies chosen to implement our PIML impact detector are described as well as the real-world test comparing a fully manual and automated workflow.

### 2.1 MiG 2.0 Protocol (field data)

We have developed an instrumented mouthguard, the MiG2.0, for measuring linear and angular head kinematics during an impact. The device is capable of sensing six degrees of freedom (6DOF) kinematics via a triaxial accelerometer (H3LIS331DL, ST Microelectronics, Geneva, Switzerland) with maximum range of ±400 g, and a triaxial gyroscope (ITG-3701, InvenSense Inc., San Jose, CA, US) with maximum range of ±4000 degrees/sec. Linear acceleration is acquired at 1000 Hz, while angular velocity is acquired at 8000 Hz. The sensor is triggered once the magnitude of linear acceleration in xyz direction (Fig.1.c) reaches 10g, and the data between the pre-trigger time of -50 ms and post-trigger time of 150ms were saved [42]. Data is temporarily stored in an on-board buffer, which allows the acquisition of pre- and post-trigger data associated with an event. When the linear acceleration threshold value set by the user is crossed at any of the linear acceleration components, an event is written to the on-board flash memory chip. Trigger threshold, pre-trigger time, post-trigger time, and sample rate of the inertial measurement units (IMUs) are user adjustable. The accuracy of all kinematic data for the MiG2.0 has been previously validated [8]

Prior to deploying the MiG2.0, research staff obtained informed consent from all participants. Participants over the age of 18 provided their own informed consent. Participants under the age of 18 had a legally authorized representatives provide consent and then provided their own assent. All consents and assents were written and followed the Stanford Institutional Review Board process. We deployed the MiG2.0 during practices and games to one college football team, Stanford University, and three local Bay Area high schools. We video analyzed the data from the collegiate players (n=12), including 5 outside linebackers, 1 inside linebacker, 1 running back, 2 wide receivers, 1 defensive end, 1 offensive tackle, and 1 center, over the course of 17 practice and game days. We also video analyzed the data from the high school players (n=49) over the course of 17 practice and game days. The high schoolers typically play multiple positions, unlike collegiate players, so our data sample included almost all football positions, except a kicker. The collegiate football games followed the National Collegiate Athletic Association (NCAA) Football Rules, and the practices were typically 2 hours in duration. The high school games and practices were also approximately 2 hours in duration, in accordance with the California Interscholastic Federation (CIF). The human subject protocols were approved by the Stanford Administrative Panels for the Protection of Human Subjects (IRB-45932 and IRB-46537). We conducted data collection and performed video analysis in accordance with the institutional review boards' guidelines and regulations. In total 1024 confirmed true impacts and 10990 confirmed false impacts were collected and processed for use in building the impact detector.

## 2.2 Finite Element Model

To generate the synthetic data, a Finite Element (FE) head-neck model was developed to simulate impacts to a digital head on the standard NFL linear pneumatic ram testbed (25,26). This FE model was developed by modifying the existing open-source, experimentally validated FE model of the standard linear impactor helmet test apparatus created by Biomechanics Consulting and Research, LLC (Biocore) [1,17,18,39,43,44]. The FE model used in this study was developed by removing the helmet parts from the original open-source FE model [43]. The modeling and simulations were conducted using LS-Dyna explicit FE software. The FE model consisted of a Hybrid III 50th percentile male anthropomorphic head-neck model and an impactor. The impactor consisted of a nylon spherical cap (i.e., impactor face), vinyl nitrile foam, circular metal backing plate, and a ram which was constrained to pure translational motion [43]. The total mass of the impactor was 15.6 kg. The head-neck model was considered to be mounted to a sliding carriage with a mass of 17.7 kg which was free to slide in the x-direction. To study head kinematic under impact loading, different impact conditions were simulated by changing initial impact configurations. The initial impact configurations were determined by head orientation angles $\alpha$ and $\beta$, head vertical position z, and initial impact velocity $v_0$ (see Figure 1). The head-neck model was symmetric about the sagittal plane and thus the impact was only applied to one side of the head sagittal plane, i.e., $-180° < \alpha < 0$ with increments of 30°. At each angle $\alpha$, different head orientations were created by considering $-30° < \beta < 30°$ with increments of 10°. Due to the asymmetry of the head about coronal and horizontal planes, at each angle $\alpha$, the variation of head vertical position z was determined through a trial-and-error process to create an even distribution of impact locations on the head and to also avoid impacts to the neck. An initial ram travel distance was considered by creating a translational offset (5 to 20 mm) between the head and the impactor. Three different initial impact velocities were considered (2 m/s, 5 m/s, and 8 m/s) to represent low-, medium- and high-energy impacts, respectively. The variation of initial impact parameters resulted in 1065 different initial impact configurations with an approximately even distribution of impact locations on the head. The head impact FE models with different initial configurations were simulated on a Linux platform using LS-Dyna version R10.1.0, single precision, with shared memory processing (SMP). The LS-PrePost was used to develop the FE models and for pre- and post-processing. The duration of the simulations was 150 ms with 0.01 ms sampling-step. The output of the FE simulations was the kinematics of the head center of gravity (CG), i.e., linear and rotational acceleration, velocity, and displacement. The head kinematics were obtained in the head coordinate system shown in Figure 1.

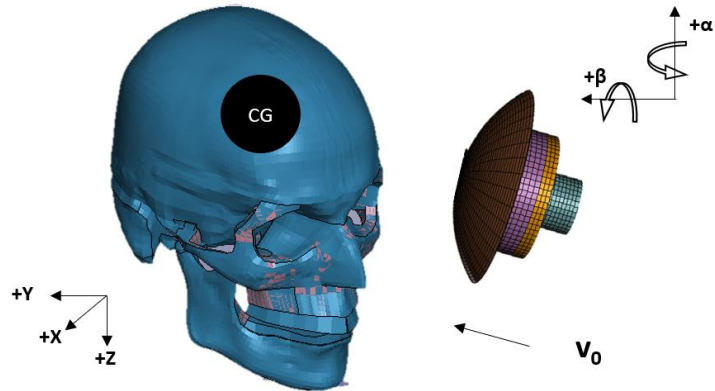

*Figure 1 - Head impact FE model with a representative initial configuration and coordinate system defined according to the Society of Automotive Engineers J211/1 standard.*

## 2.3 Impact Detector Training

With both on field data gathered and verified with video analysis and the simulated on field data generated by the Finite Element head-neck model, a strategy was developed to test the efficacy of introducing synthetic data to the training of the impact detector.

The dataset consisted of comma separated value (csv) files that contain the time series signals of recorded mouthguard kinematic data. This included 6 variables: the x,y,z spatial components of linear acceleration as well as the three components of angular acceleration. These signals were 200ms in length with the triggering point placed 50ms into the signal and 150ms before the end. Data from the FE model was also produced to match the style output from the mouthguard. Based on the work by Domel et al. in (13) a deep 1D convolutional neural network architecture was constructed using the same number and type of layers and neurons as used in their work. This consisted of 2x [1D convolution layer (64 (for the first) and 128 (for the second) filters, 5 × 1 (for the first) and 10 × 1 (for the second) filter sizes, ReLU activation), max pooling (pool = 2), batch normalization layer, dropout layer 40%], a reshaped 2D combination of layer one and layer 2 to create a 2D matrix of combined layers, 2D convolution (64 filters, 3 × 15 patch, 3 × 1 stride size, ReLU activation), global average pooling, batch normalization layer, dropout layer with 40%, output layer with Softmax activation. To ascertain which data configuration would result in the best performance, the following training strategies were implemented both with and without an augmentation strategy to reach a balanced training dataset:

1) Train on just the sensor data
2) Train on field data and 25% synthetic data
3) Train on field data and 100% synthetic data
4) Pre-train the network on synthetic data and re-train on field data

In all cases, field data was separated into training, validation, and testing with a 70%-15%-15% split. No synthetic data was used for validation or testing as the goal was to improve the algorithm's ability to detect real-world cases.

In total, 1024 confirmed true impacts and 10,990 confirmed false impacts were collected from video verification of real-world sensor data. Due to the complexity of the signals and the imbalance in the data, an augmentation approach was used to increase both the total number of samples, and to improve the balance of true and false samples. Firstly, to increase the number of signals used to train with, the signal of each kinematic variable was shifted along the time axis by a set unit of milliseconds. The total number of augmented signals from one original sample was a controlled variable. In this study, a maximum of 5 augmented samples were obtained, meaning that the signal was offset by 1-5ms for each collected signal. This allowed the largest training set of field data to consist of a total of approximately 5100 true impact signals and approximately 55,000 false impact signals.

Additionally, as there existed an imbalance of 1:10 for the true to false impact class ratio, further modifications were utilized. By calculating the different weighting for each class, the loss function was augmented during training, to weigh the improvements between the two classes more evenly. For strategies 1-5, the datasets were prepared in accordance with the relative inclusions of synthetic data, and then the model was trained on the training and validation data before being used to analyze the same test dataset. For the pre-training case, all of the synthetic data was augmented by a factor of 5 and the model trained on a combination of synthetic impacts and verified false impacts from field data (this was done given the vast amount of false impact data available). Once the pre-training had converged, a further set of training epochs were initiated, this time with the model using only the on-field data, with the weights and biases beginning from the last state of being pre-trained. The motivation for this pre-training technique will be discussed later.

To compare the different strategies, the following performance variables were used to analyze and compare their performance:
- Positive predictive value (PPV) - Proportion of predicted true results that are true.
- Negative predictive value (NPV) - Proportion of predicted false results that are false.
- F-Score

The F-score is useful to compare the effectiveness of a binary classifier as it represents the harmonic mean of the Recall and Precision of a model.

$$F_X = (1 + X^2)(\frac{Precision \cdot Recall}{(1 + X^2) \cdot Precision + Recall})$$

An $F_1$ score (X = 1) is a balanced weighting between recall and precision while an $F_2$ score (X = 2) places more importance on precision than recall. This $F_2$ is important in this case as we wanted to ensure the minimum amount of true impacts were classified incorrectly.

## 2.4 Manual vs. Auto workflow

With the overall objective of the development of this new impact detection algorithm being a potential replacement for traditional video analysis, a real-world test was conducted to compare both workflows in terms of efficiency and accuracy. Six participants were chosen, and their data followed through the Spring of 2021. Data was collected as per the same methodology found in [13]. Mouthguard data was then filtered for 6 participants. Of these participants, the data collected indicated 88 total recorded impacts from their mouthguards. These impacts were first each verified by video analysis to determine if the events were positive or negative impacts. Following this, these 88 events were passed through the impact detection algorithm to label each event as positive or negative for impacts. To compare approaches, the time to complete each workflow, and the percentage of true and false impacts correctly inferred by the algorithm (assuming the video analysis was the ground truth) was recorded. To be considered a reasonable replacement for video analysis, a minimum F2 score of 0.90 was deemed appropriate for the impact detector to achieve.

## 3. Results

The results of the six strategies, and the direct comparison of manual video analysis and a fully automated workflow are presented in this section.

### 3.1 Training on a mixture of on field and synthetic data

As the most naive and initial approach to obtain a base level of performance, the initial results consisted of the training of the network directly on the data gathered and verified from on field. This was done by training on 1024 true impacts and 10990 false impacts. The NPV and PPV values for this trained network were 0.69 and 0.97, respectively. Including synthetic data to the training dataset improved the network performance to a maximum of 0.72 and 1.0 for the NPV, and PPV, respectively.
Figure 2 shows these results as a function of included synthetic data for the unaugmented case. Adding augmentation to the training datasets meant an increase by a factor of 5 of the samples. By adding synthetic data in increments of 25% of the true impact sample size, the maximum network performance reached 0.76 and 0.97 for the NPV and PPV, respectively, when 25% synthetic data was added to the training dataset. The results as a function of included synthetic data with augmentation is shown in Figure 3.
Finally, to address the imbalance in the dataset, the number of false impact results was reduced to match the number of true impact results, with augmentation at a factor of 2. Peak network performance was found at 100% additional synthetic data with NPV and PPV of 0.87 and 0.86, respectively. The performance of this balanced training data approach is shown in Figure 4.

### 3.2 Pre-Training and Synthetic Data

The pre-trained network which was initially trained for 30 epochs on only the full dataset of synthetic true impacts, and equal amount of verified false impacts, and then further trained for another 30 epochs on a new training dataset with a balanced dataset of video verified true and false impacts, produced NPV and PPV of 0.89 and 0.86, respectively. The pre-trained network and the combined synthetic network performance values are shown in Figure 2 - Figure 5.

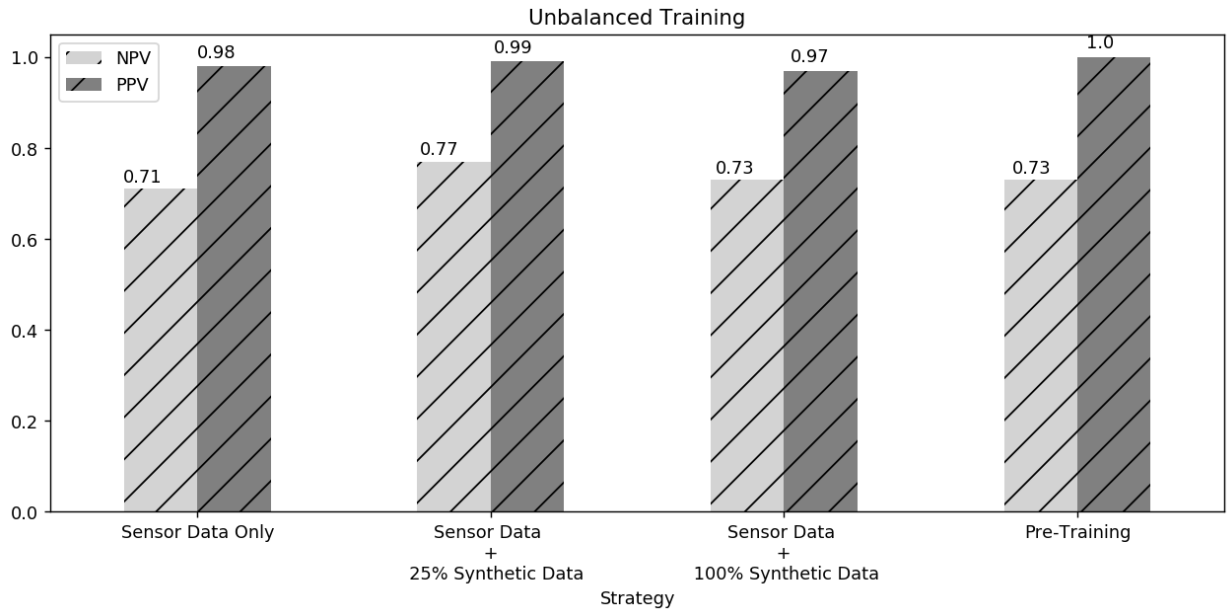

*Figure 2 - Comparison of different training strategies on the NPV and PPV values of the test performance as a function of added synthetic data. In these training sessions, no augmentation was performed and thus the datasets had an imbalance of true and false of 1:10.*

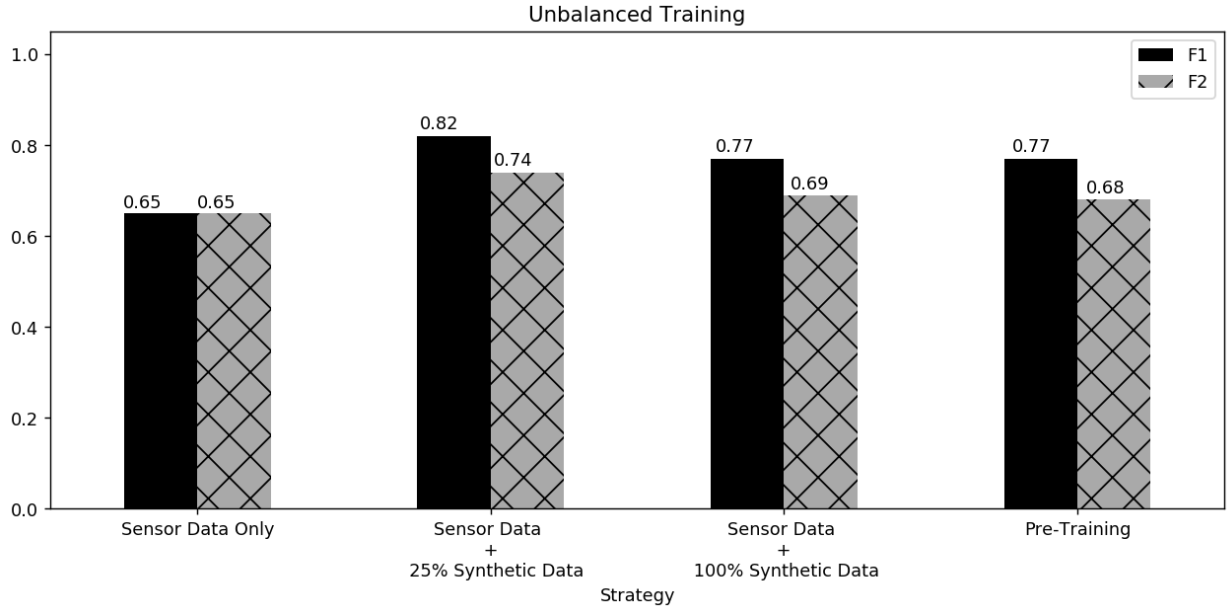

*Figure 3 - Comparison of different training strategies on the $F_1$ and $F_2$ scores of the test performance as a function of added synthetic data. In these training sessions, no augmentation was performed and thus the datasets had an imbalance of true and false of 1:10*

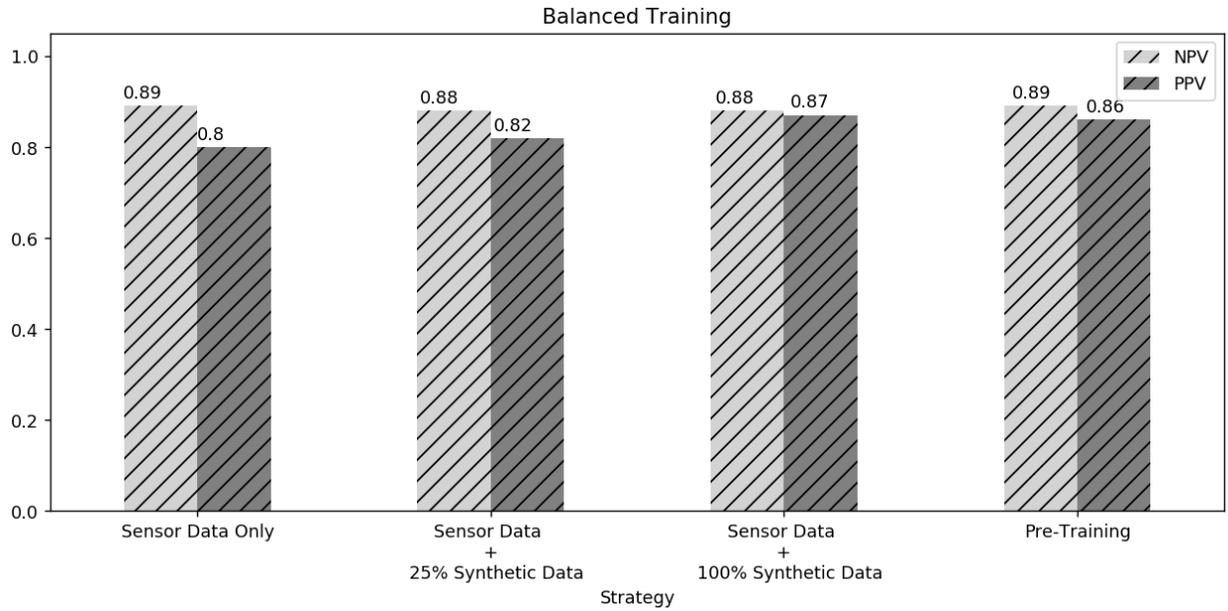

*Figure 4 - Comparison of different training strategies on the NPV and PPV values of the test performance as a function of added synthetic data. In these training sessions, additional augmentation was performed so that the dataset had a balanced amount of true and false samples.*

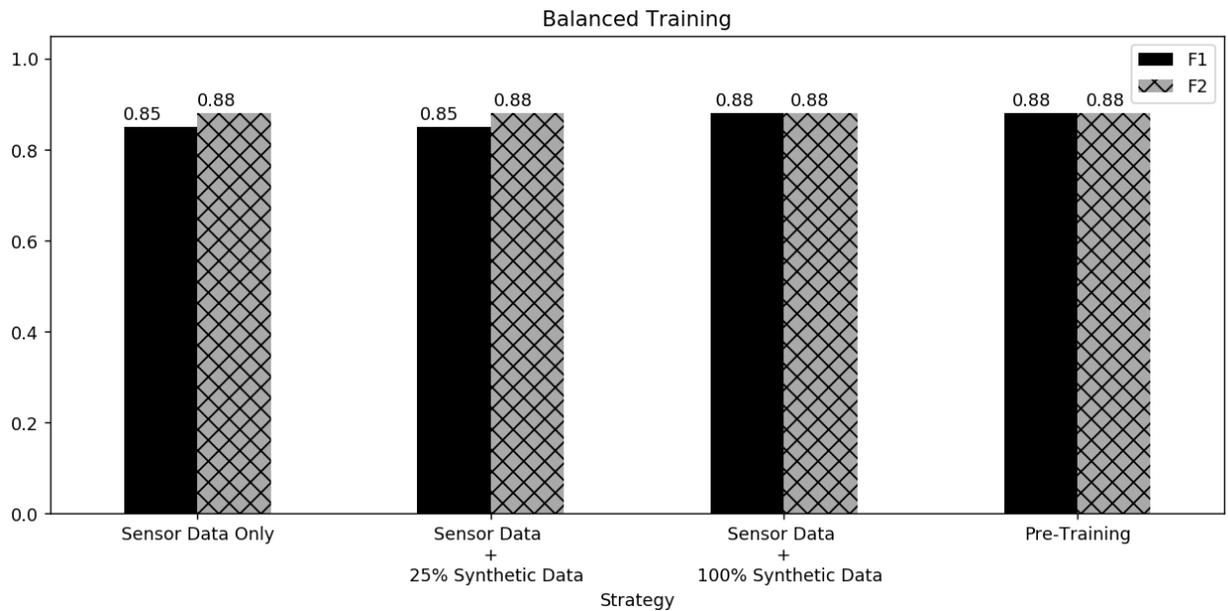

*Figure 5 - Comparison of different training strategies on the $F_1$ and $F_2$ scores of the test performance as a function of added synthetic data. In these training sessions, additional augmentation was performed so that the dataset had a balanced amount of true and false samples.*

### 3.3 Manual and fully automated workflow comparisons

To compare the accuracy and efficiency of a manual workflow to a fully automated one, the time to train a video analyst, and the time taken to review a set number of videos for confirmed true or false impacts was compared to the time taken to train a fully automated workflow and its accuracy to detect the same true and false impacts. **Error! Reference source not found.** shows the breakdown of these numbers as a direct comparison.

|  | Manual Video Analysis | Impact Detection Algorithm |
|---|---|---|
| # Hours Processing | ~12 (20 events/hour) | ~0 (>1e6 events/hour) |
| # Hours Training | 2 | 5 |
| True Impacts* Found | 61 | 61 |
| False Impacts* Found | 27 | 20 |
| False Positives Found |  | 7 |
| False Negatives Found |  | 0 |

Table 1 - Results from the real-world test comparing a traditional manual video analysis of Football played in a season of spring football and the performance of the PIML-based impact detection system. *All impacts recorded by the mouthguard were triggered by an event of 10 G impact or higher.

### 3.4 Comparison to other Impact Detectors

While it is difficult to compare different impact detection algorithms published in the literature due to different training and testing datasets, one important measure that represents the accuracy rate of these modes is the F-score. The work presented here was compared to published results of impact detectors, also for American football with mouthguard data. Table 1 shows the relative performances of these models based on their $F_1$ and $F_2$ scores.

| Model | Model Architecture | $F_1$ Score | $F_2$ Score |
|---|---|---|---|
| Domel et al. 2020 (13) | Convolutional Neural Network | 0.79 | 0.75 |
| Gabler et al. 2020 (14) | AdaBoost CART Decision Tree | 0.89 | 0.84 |
| Wu et al. 2018 (16) | Support Vector Machine | 0.90 | 0.88 |
| MiGNet 2.0 (this work) | PIML Convolutional Neural Network | 0.95 | 0.98 |

Table 1 - Comparison of the $F_1$ and $F_2$ scores of different impact detection models presented in the literature and the PIML-based impact detector presented in this work.

### Discussion

In this work a new impact detection algorithm was created, based on physics-informed machine learning (PIML). This algorithm was trained on kinematic data of athletes playing American Football and high school and collegiate practice and game sessions. Additionally, Finite Element simulations of head-neck impacts were constructed to supply the dataset with synthetic data in an attempt to compensate for the largely imbalanced composition of true and false impacts. As the instrumented mouthguards used to record kinematic data are unable to differentiate other body motions that can mimic head impacts during gameplay, an impact detector can be used on the data these mouthguards record to intelligently discern true impacts from false ones. Current practices rely on manual video analysis using a synchronized dataset and video footage library in order to confirm impacts are real or false. This process is necessary to ensure that appropriate attention is given to those who receive sufficient impacts, but at the cost of many manual hours of analysis. Previous attempts to circumvent this manual process have been suggested based on machine learning tools such as support vector machines or deep learning neural networks, but the imbalance in the data and the access to large amounts of true impacts have prevented these automated workflows from being fully adopted. PIML represents an attempt to meld the fields of computational mechanics and machine learning to take advantage of domain knowledge and data science advances. In this current work, as true impacts are rare compared to false impacts at a rate of 10:1, a Finite Element head-neck model, meant to represent a football player, was subjected to a wide range of typical impact loads. These simulations resulted in a new source of synthetic data, with a potentially limitless generating ability. Combining these synthetic results with the manually verified data enabled an investigation of the efficacy of enriching this training data set on a deep neural network impact detector. The overall aim of this approach being to present a realistic alternative to manual video analysis for impact detection.

Figure 2 and Figure 4 show the results of the strategies used to compare impact detectors trained on various amounts of synthetic data and the application of augmentation techniques to increase the dataset size. Encouragingly, as more synthetic data was included into the training dataset, the performance of the detector improved relative to a common test set. This test set only included real-world impacts to ensure that the algorithm was improving its ability to detect real events, not simulated ones. Despite a relatively simple impact simulation example and a modest increase in the number of true impacts with the increase in synthetic data, the improvements show a marked increase over the naive case. The best performing configuration in the first 5 strategies attempted showed an NPV of 0.88 and a PPV of 0.87. While the PPV was higher in other cases, due to the importance of ensuring that a true impact was not mistaken for a false result, i.e. tolerating more false positives than false negatives, the combination of equal amounts of true impacts from on field and simulated data yielded the best results. To better see the effect of this, F-scores of the different strategies were also calculated and shown in Figure 3 and Figure 5. These results show how the PPV results can be misleading and result from an unbalanced dataset. By adding synthetic data to balance the dataset, both $F_1$ and $F_2$ scores improved. This was promising as simulated impacts could be generated at will, and conceivably any number of impacts could be generated to match the number of verified data samples.

The sixth strategy performed in this work revolved around the concept of pre-training. In pre-training, a network is trained on synthetic or representative data initially. This serves to guide the loss function toward a global minimum, using as much data as possible, before using what could be scarce amounts of real data. Pre-training in the context of PIML is a useful strategy as data collection can be both time consuming and be of low yield if the data being gathered is precarious or rare. The results shown in Figure 5 indicate that a similarly performant impact detector can be developed using pre-training which could be valuable for other cases where data is rare or difficult to collect on the field. This could also be used as a form of transfer learning when attempting to use one impact detector built from one sport to another sport type.

The final results of this work focused on a real-world test to directly compare this impact detector to a traditional video analysis approach whereby the mouthguard events are each checked by a human via the recorded video footage. Using the most performant impact detector, athletes' data was analyzed from the Spring of 2021 and Table 1 shows the results from the video analysis and the impact detector. In total, 88 impacts were analyzed resulting in 61 true and 27 false video verified impacts. The impact detector identified 68 true and 20 false impacts, with NPV and PPV, of 1.00 and 0.90, respectively. With only 7 misclassified impacts this model achieved F1 and F2 scores of 0.98 and 0.95, respectively. For this real-world test these results are encouraging and are well above the authors minimal requirements to use this impact detector in an automated system where no video analysis is required. In addition, given the time to train and run inferences for this impact detector was several orders of magnitude shorter (1 event/min for manual analysis vs. 100,000 events/min for the algorithm) than that to train a video analyst and have them review all of the game footage in this example, the automated impact detector is also a valuable addition to this project in terms of time efficiency and scalability.

While the results of this study are promising, there remain issues with the dataset that persist. For example, the vast majority of impacts recorded on the mouthguard occurred offscreen from the video and so these events are labeled as unknown and hence not used in the analysis. Given that the players are almost never involved in actual activity offscreen, and hence these are likely from chewing or mishandling of the mouthguard, this presents a problem. If the impact detector is used completely without human oversight, these offscreen events, which are likely device misuse, would be incorrectly flagged. It is possible that an additional automated workflow could be devised to check for the number of the player being onscreen at the time of the event, but this was not attempted in this work. Also, while the FE model was used to great effect in generating synthetic data, it is likely that the scope of impact types that can occur is much larger than this modest setup can reproduce. This would require a more complex model with more motor functional control to generate more true to life impacts. Finally, this model was only trained and tested on the same athletes in the same sport. It is possible that the model would not extend to all other athletes playing American football, or the same people playing another sport with potential head impacts.

Moving forward though, this PIML approach opens new opportunities for research as this allows new models of body dynamics in different sports to be computational modelled to generated the same kind of synthetic data so that larger datasets can be produced and trained on so that impact detectors can be made for a variety of sports. These models do not need to be complex FE models of human movement, but will need to provide data that is sufficiently complex to mimic the data provided from wearable sensors. New techniques such as generative adversarial networks could also provide a source of synthetic data by learning from sensor data. As Table 2 indicates, this approach appears to surpass previous attempts in the field and can be abstracted to other sports. With such encouraging results in this study, more application of this impact detector should be undertaken to build more confidence in using an automated system to detect these impacts and improve the health monitoring of athletes. A potential new workflow could be to utilize the impact detector as a primary evaluation tool and periodically select random samples of data to verify the continued performance of the model. Retraining of the neural network format is straightforward and thus further improvements can be made continuously as needed.

## Conclusion

A physics-informed machine learning impact detector was created to analyze kinematic data from an instrumented mouthguard. This was used to monitor player impacts that could produce mild traumatic brain injury. While a combination of video analysis and sensor data is often used to detect true impacts, due to the nature of manual video analysis and the. Here we show that by simulating head impacts numerically using a standard Finite Element head-neck model, a large dataset of synthetic impacts can be created to augment the gathered, verified, impact data from mouthguards. This combined physics-informed machine learning impact detector reported improved performance on test datasets compared to traditional impact detectors with negative predictive value, positive predictive value and accuracies of 88%, 87% and 88% respectively. After a test on new data this model reported the best results to date for an impact detection algorithm for American Football, achieving an $F_1$ score of 0.95. In addition, this new model was able to accurately detect true and false impacts from a new dataset at a rate of 90% and 100% relative to a purely manual video analysis workflow. Saving over 12 hours of manual video analysis for a modest dataset, at an overall accuracy of 92%, these results indicate that this model could be used in place of, or alongside, traditional video analysis to allow for larger scale and more efficient impact detection in sports such as American Football.